\documentstyle[psfig,preprint,aps]{revtex}

\def\nps#1#2#3{           Nucl. Phys. B (Proc. Suppl.) {\bf #1}, #3 (19#2)} 
\def\np#1#2#3{            Nucl. Phys. {\bf #1}, #3 (19#2)}
\def\pl#1#2#3{            Phys. Lett. {\bf #1}, #3 (19#2)}
\def\pr#1#2#3{            Phys. Rev. {\bf #1}, #3 (19#2)}

\def\sjnp#1#2#3{          Sov. J. Nucl. Phys. {\bf #1}, #3 (19#2)}
\def\jetp#1#2#3{          Sov. Phys. JETP {\bf #1}, #3 (19#2)}

\def\ppnp#1#2#3{            Prog. Part. Nucl. Phys. {\bf #1}, #3 (19#2)}
\def\yf#1#2#3{            Yad. Fiz. {\bf #1}, #3 (19#2)}
\def\eq#1{{eq.~(\ref{#1})}}
\def\Eq#1{{Eq.~(\ref{#1})}}

\def\Frac#1#2{\frac{\displaystyle{#1}}{\displaystyle{#2}}}
\def\lsim{\raise0.3ex\hbox{$\;<$\kern-0.75em\raise-1.1ex\hbox{$\sim\;$}}}
\def\gsim{\raise0.3ex\hbox{$\;>$\kern-0.75em\raise-1.1ex\hbox{$\sim\;$}}}
\begin{document}
\title{Neutrino magnetic moments and low-energy
solar neutrino-electron scattering experiments}
\author{S. Pastor~$^a$\footnote{ E-mail: sergio@flamenco.ific.uv.es},
J. Segura~$^b$\footnote{ E-mail: javier.segura@umh.es},
V.B. Semikoz~$^a$
\footnote{ E-mail: semikoz@izmiran.rssi.ru~
On leave from {\sl Institute of Terrestrial Magnetism, the Ionosphere
and Radio Wave Propagation of the Russian Academy of Sciences,
IZMIRAN, Troitsk, Moscow region, 142092, Russia}.}
J.W.F. Valle~$^a$
\footnote{ E-mail valle@flamenco.ific.uv.es}}
\address{$^a$ Instituto de F\'{\i}sica Corpuscular - C.S.I.C.\\
Departament de F\'{\i}sica Te\`orica, Universitat de Val\`encia\\
46100 Burjassot, Val\`encia, SPAIN \\
http://neutrinos.uv.es}
\address{$^b$ Instituto de Bioingenier\'{\i}a \& \\
Departamento de Estad\'{\i}stica
y Matem\'atica Aplicada\\
Universidad Miguel Hern\'andez, Edificio La Galia\\
03206 Elche, Alicante, SPAIN}

\maketitle
 
\begin{abstract}

The scattering of solar neutrinos on electrons is sensitive to the
neutrino magnetic moments through an interference of electromagnetic
and weak amplitudes in the cross section.  We show that future
low-energy solar neutrino experiments with good angular resolution can
be sensitive to the resulting azimuthal asymmetries in event number
and should provide useful information on non-standard neutrino
properties such as magnetic moments. We compare asymmetries expected
at Hellaz (mainly pp neutrinos) with those at the Kamiokande and
Super-Kamiokande experiments (Boron neutrinos), both for the case of
Dirac and Majorana neutrinos and discuss the advantages of {\sl low
energy experiments}. Potentially interesting information on the solar
magnetic fields may be accessible.

\end{abstract}

\newpage

\section{Introduction}

Most non-standard properties of neutrinos arise from non-zero masses
\cite{fae,revnu}. Among these electro-magnetic dipole moments play an
important role \cite{MoPal}.  Here we are concerned with a particular
effect in neutrino-electron scattering for neutrinos from the Sun
which possess a Dirac magnetic moment \cite{VVO} or transition
magnetic moments \cite{BFD} in the case of Majorana neutrinos. The
latter is especially interesting first of all because it is more
fundamental theoretically, and because Majorana neutrinos are the ones
which arise in most extensions of the Standard Model. Moreover, the
effects of Majorana transition moments can be resonantly enhanced when
neutrinos propagate in media \cite{RFSP} such as the Sun, providing
one of the attractive solutions to the solar neutrino problem
\cite{akhmedov97}. Another practical advantage in favour of Majorana
transition moments is that, in contrast to Dirac-type magnetic
moments, these are substantially less stringently constrained by
astrophysics \cite{Raffelt}.

For {\sl pure left-handed neutrinos} the weak interaction and the
electro-magnetic interaction amplitudes on electrons do not interfere,
since the weak interaction preserves neutrino helicity while the
electro-magnetic does not. As a result the cross section depends
quadratically on $\mu_\nu$.
 
However, if there exists a process capable of converting part of the
initially fully polarized $\nu_e$'s, then an {\sl interference term}
arises, proportional to $\mu_\nu$, as pointed out e.g. in
ref. \cite{Barbieri}.  This term depends on the angle between the
component of the neutrino spin transverse to its momentum and the
momentum of the outgoing recoil electron.  Therefore the event count
rates expected in an experiment would exhibit an {\sl asymmetry} with
respect to the above defined angle.  Such asymmetry would not show up
in earth-bound laboratory experiments even with stronger magnetic
fields, since the helicity-flip could be caused only by the presence
of a neutrino mass and is therefore negligible \cite{grimus}. However,
in the solar convective zone one may find a magnetic field extended
over a tenth or so of the solar radius and, most importantly, the
neutrino depolarization could be resonant in the Sun. Even if the Sun
possesses only a relatively modest large-scale magnetic field
$B_{\perp} \sim 10^4$ G in the convective region ($L\sim L_{conv}
\simeq 3\times 10^{10}$ cm), and for a neutrino magnetic moment of the
order $10^{-11} \mu_B$ such a spin-flip process may take place with
sizeable rates, since in such a case one has $\mu_{\nu} B_{\perp}L
\sim 1 $.

Barbieri and Fiorentini considered \cite{Barbieri} the conversions
$\nu_{eL} \to \nu_{eR}$ in the Sun as a result of the spin-flip by a
toroidal magnetic field in the convective zone. They showed that the
azimuthal asymmetry could be observable in a real time solar
$^8$B-neutrino experiment and as large as 20\% for an electron kinetic
energy threshold of $W_e=5$ MeV.  They chose a fixed $\nu_e$ survival
probability $P_e=1/3$ (as suggested at that time by the Homestake
experiment) and the maximal Dirac magnetic moment allowed by
laboratory experiments, $\mu_\nu\simeq 10^{-10}\mu_B$.

On the other hand, Vogel and Engel \cite{Vogel} emphasized that if an
asymmetry in the scattering of solar neutrinos exists, recoil
electrons will be emitted copiously along the direction of the
neutrino polarization in the plane orthogonal to the neutrino
momentum. They calculated the asymmetry expected for solar $^8$B
neutrinos with $\mu_\nu=10^{-10} \mu_B$ and concluded that it would be
difficult to detect because of the poor angular resolution of the
experiments. Moreover, as we will see later, both \cite{Barbieri} and
\cite{Vogel} {\sl overestimated} the asymmetry. Thus their
calculations are not accurate. 

In this paper we correct results for the asymmetry in the case
presented by \cite{Barbieri} and \cite{Vogel} for high energy $^8$B
neutrinos.  In addition we compare them with the expected asymmetry in
the case of a Majorana transition magnetic moment of the same
magnitude. More importantly, we show the sensitivity of planned solar
neutrino experiments in the {\sl low energy region} ($\omega \lsim 1$
MeV) to the azimuthal asymmetries that are expected in the recoil
electron event rates, arising from the above electro-weak interference
term.  We calculate the asymmetry for the low-energy $pp$-neutrinos
fixing the survival probability at $P_e=0.5$. This gives the maximum
expected asymmetry and seems phenomenologically reasonable in order to
convert the initial solar $\nu_{eL}$'s via the Resonant Spin-Flavour
Precession (RSFP) scenario.  In particular we calculate the asymmetry
that could be observed in the azimuthal distribution of events in an
experiment like the proposed Hellaz \cite{Hellaz}, sensitive to the
fundamental $pp$ neutrinos from the Sun. The Multi-Wire-Chamber in
Hellaz should measure both the recoil electron energy $T$ and the
recoil electron scattering angle $\theta$ with good precision.
Moreover Hellaz should be sensitive to the azimuthal angle $\phi$,
measuring the number of events in $\phi$--bins.  We discuss the
sensitivity of the Hellaz experiment for probing $\mu_\nu$ and compare
it with planned accelerator experiments.  In particular there are very
interesting new projects, such as the future ITEP-Minnesota
experiment, where they plan to search $\mu_{\nu}/\mu_B$ down to
$3\cdot 10^{-11}$ with reactor anti-neutrinos \cite{Voloshin}, and the
LAMA experiment, which will use a powerful isotope neutrino source
\cite{Bernabei}.

Finally, we also refine our calculations of the azimuthal asymmetry
expected for $pp$ neutrinos at Hellaz using a realistic energy-dependent
conversion probability $P_e$ based on a simple model for resonant spin
flip conversions in the Sun.

\section{Neutrino-electron Scattering Cross sections}

We consider the neutrino--electron scattering process $\nu_e(k_1)
+e(p_1) \to \nu_e(k_2) +e(p_2)$ when the initial flux of neutrinos is
not completely polarized, as a result of conversions induced by non-zero
transition magnetic moments in the Sun. In other words, our flux is a
mixture of both the original left-handed solar neutrinos with the
converted ones (right-handed).  We consider both conventional
Dirac-type magnetic moments as well as Majorana transition
moments. The differential cross section can be written as a sum of
three terms, (for details, see for instance \cite{Sem97})
\begin{equation}
\frac{d\sigma}{dTd\phi} = \left
(\frac{d\sigma}{dTd\phi}\right )_{weak} +
 \left (\frac{d\sigma}{dTd\phi}\right )_{em} +
\left (\frac{d\sigma}{dTd\phi}\right )_{int},
\label{totalS}
\end{equation}
where $T$ is the recoil energy of electrons and $\phi$ the azimuthal
angle defined in figure \ref{axis}. 

Let us first assume that only the Dirac $\nu_e$ magnetic moment
exists, $\mu_{\nu_e}$. For ultra-relativistic neutrinos the
expressions for the weak and electro-magnetic terms are the following
\begin{equation}
\left (\frac{d\sigma}{dTd\phi}\right )_{weak} =
P_e \frac{G_F^2m_e}{\pi^2}
\left [ g_{eL}^2 + g_R^2\left (1 - \frac{T}{\omega}\right )^2 -
\frac{m_eT}{\omega^2}g_{eL}g_R\right ]
\label{weakDirac1}
\end{equation}
\begin{equation}
\left
 (\frac{d\sigma}{dTd\phi}\right )_{em} = \frac{\alpha^2 }{2m_e^2}
\left (\frac{\mu_{\nu_e}}{\mu_B}\right
)^2\left [\frac{1}{T} - \frac{1}{\omega}\right ]~.
\label{em}
\end{equation}
Here $P_e$ is the survival probability of the initial $\nu_{eL}$,
$\omega$ is the incoming neutrino energy, and $g_{eL} = \sin^2
\theta_W + 0.5$ and $g_R = \sin^2 \theta_W$ are the constants of the
Standard Model with $\sin^2\theta_W \simeq 0.23$.

In such a case there is an {\sl interference term} proportional to
$\mu_\nu$ \cite{Barbieri},
\begin{equation}
\left (\frac{d\sigma}{dTd\phi}\right )_{int} =
-\frac{\alpha G_F}{2\sqrt{2}\pi m_e T} 
\left(\frac{\mu_{\nu_e}}{\mu_B}\right )\vec{p}_2 \cdot \vec{A}_{D} (T,\omega)
\label{diracdiag}
\end{equation}
where
\begin{equation}
\vec{A}_{D} (T,\omega) \equiv \left[g_{eL} + 
g_R\left (1 - \frac{T}{\omega}\right )\right ]
\vec{\xi}_\perp  
\label{add}
\end{equation}
Here $\vec{\xi}_\perp$ is the transverse component of the neutrino
polarization spin vector with respect to its momentum. It is a
function of the $\nu_{eL}$ survival probability, $\mid
\vec{\xi}_{\perp} \mid = 2\sqrt{P_e (1 - P_e)}$. This interference
term depends on the angle between $\vec{\xi}_\perp$ and the momentum
of the outgoing electron $\vec{p}_2$.  \Eq{diracdiag} may be written
as a function of the azimuthal angle $\phi$ (as defined in figure
\ref{axis}) using
\begin{equation}
\vec{p}_2\cdot\vec{\xi}_\perp = \mid \vec{p}_2 \mid 
\sin \theta \mid \vec{\xi}_\perp \mid \cos \phi = \sqrt{2m_e
T\left (1 - \frac{T}{T_{max}}\right )} \mid \vec{\xi}_\perp \mid \cos
\phi
\label{angles}
\end{equation}
where $T_{max} = 2\omega^2/(m_e + 2\omega)$ is the maximum electron
recoil energy.

Let us explain here why the results of \cite{Barbieri} and
\cite{Vogel} for Dirac neutrinos are not correct. First, the weak term
in eq.~(11a) of Barbieri \& Fiorentini is a factor 2 less than our
\eq{weakDirac1}, while the interference term coincides with ours.  On
the other hand Vogel \& Engel used an interference term (their
eq.~(A9)) which is a factor 2 bigger than our \eq{diracdiag}. Thus
both overestimated the asymmetry by an approximate factor two. This
also agrees with a recent calculation in reference \cite{Gaida}.

If neutrinos are Majorana particles they can only possess a transition
magnetic moment $\mu_{12}$. For simplicity we assume the case of CP
conservation.  For definiteness, moreover, we consider the case of two
neutrino species, $\nu_e$ and $\nu_\mu$, with positive relative
CP-parity \cite{BFD}.  The three terms of the differential cross
section will include an electro-magnetic term (same as
\eq{em} with $\mu\to  \mu_{12}$), a weak term,
$$
\left (\frac{d\sigma}{dTd\phi}\right )_{weak} =
\frac{G_F^2m_e}{\pi^2}\Biggl \{
 P_e\left [ g_{eL}^2 + g_R^2\left (1 - \frac{T}{\omega}\right )^2 -
\frac{m_eT}{\omega^2}g_{eL}g_R\right ] + 
$$
\begin{equation}
+ (1 - P_e)\left [g_R^2 +
g_{\mu L}^2\left (1 - \frac{T}{\omega}\right )^2 -
\frac{m_eT}{\omega^2}g_{\mu L}g_R\right ]\Biggr \}~,
\label{weak1}
\end{equation}
where $g_{\mu L}=\sin^2 \theta_W - 0.5$, and an interference term
\begin{equation}
\left (\frac{d\sigma}{dTd\phi}\right )_{int} =
-\frac{\alpha G_F}{4\sqrt{2}\pi m_e T} 
\left(\frac{\mu_{12}}{\mu_B}\right )\vec{p}_2 \cdot \vec{A}_M (T,\omega)
\label{majtran}
\end{equation}
where we have defined
\begin{equation}
\vec{A}_M (T,\omega) \equiv \left[(g_{eL} + g_{\mu L} + 2g_R)
\left (2 - \frac{T}{\omega}\right )+
(g_{eL} - g_{\mu L})
\frac{T}{\omega}\right ]
\vec{\xi}_\perp^{~e \bar{\mu}}
\label{amaj}
\end{equation}
Here the mixed polarization vector is given by \cite{Sem97}
\begin{equation}
\mid \vec{\xi}_\perp^{~e\bar{\mu}} \mid = 
2 \sqrt{P_e P_{\bar{\mu}}}
\label{mixed}
\end{equation} 
where $P_{\bar{\mu}}=1-P_e$ is the $\nu_{eL} \to \bar{\nu}_{\mu R}$
conversion probability.

\section{Expected azimuthal asymmetries in Hellaz}

The relevant quantity to be measured in neutrino-electron scattering
experiments capable of measuring directionality of the outgoing
$e^{-}$ (like Hellaz) is the event number azimuthal distribution,
namely
\begin{equation}
\label{spectrum}
\frac{dN}{d\phi} = N_e \sum_{i} \Phi_{0i} 
\int^{T_{max}}_{T_{Th}}dT~
\int^{\omega_{max}}_{\omega_{min}(T)}
d \omega~ \lambda_i(\omega) \epsilon(\omega)
\frac{d \sigma}{dTd\phi} (\omega,T) 
\end{equation}
where $d \sigma/dTd\phi$ is given in \eq{totalS}, $\epsilon(\omega)$
is the efficiency of the detector (which we take as unity for energies
above the threshold, for simplicity), and $N_e$ is the number of
electrons in the fiducial volume of the detector.  The sum in the
above equation is done over the solar neutrino spectrum, where $i$
corresponds to the different reactions $i= pp$, $^7$Be, $pep$, $^8$B
$\ldots$, characterized by a differential spectrum $\lambda_i(\omega)$
and an integral flux $\Phi_{0i}$.

In the previous section we found that the azimuthal distribution of
the number of events can be written as
\begin{equation}
\Frac{dN}{d\phi}=n_{weak}+n_{em}+ n_{int} \cos\phi 
\end{equation}
where $n_{weak}$ ($n_{em}$) accounts for the weak (electro-magnetic)
contributions, while $n_{int}$ is the interference term.

The differential azimuthal asymmetry is defined as
\begin{equation}
\left.\Frac{dA}{d\phi}\right|_{\phi '}=
\Frac{\left.\Frac{dN}{d\phi}\right|_{\phi '}-
\left.\Frac{dN}{d\phi}\right|_{\phi '+\pi}}{
\left.\Frac{dN}{d\phi}\right|_{\phi '}+
\left.\Frac{dN}{d\phi}\right|_{\phi '+\pi}}=
\Frac{n_{int}}{n_{weak}+n_{em}} \cos\phi '
\end{equation}
where $\phi$ ($\phi '$) is measured with respect to the direction of
the magnetic field $\vec{B}_\odot$, which we will assume to be along the
positive $x$-axis (see fig.~\ref{axis}).

One can also define an integrated (over $\phi$) asymmetry 
\begin{equation}
{\cal A}(\phi ')=\Frac{N_2 (\phi ')-N_1 (\phi ')}{
N_2 (\phi ')+N_1 (\phi ')}~,
\end{equation}
where
\begin{equation}
N_2 (\phi ')={\displaystyle \int^{\phi '+\pi}_{\phi '}} \Frac{
dN}{d\phi}d\phi =
\pi(n_{weak}+n_{em})-2n_{int}\sin \phi '
\end{equation}
\begin{equation}
N_1 (\phi ')={\displaystyle
 \int^{\phi '+2\pi}_{\phi '+\pi}} \Frac{
dN}{d\phi}d\phi=
\pi(n_{weak}+n_{em})+2n_{int}\sin \phi '
\end{equation}
and then one gets
\begin{equation}
{\cal A}(\phi ')=-\Frac{2n_{int}}{\pi (n_{weak}+n_{em})}\sin\phi '
\label{asym4}
\end{equation}
which is directly related to the differential asymmetry by 
$${\cal A} (\phi ')= -\Frac{2}{\pi} \left.\Frac{dA}{d\phi}\right|_{\phi '}
\tan\phi '
$$
Let us define the maximum integrated asymmetry measurable by the 
experiment,
\begin{equation}
A=\Frac{2 n_{int}}{\pi (n_{weak}+n_{em})}
\end{equation}
Then
\begin{equation}
{\cal A} (\phi ')=-A \sin \phi '
\end{equation}
where $A$ is manifestly positive.

It is important to emphasize that Hellaz will be the first experiment
which is potentially sensitive to azimuthal asymmetries since the
directionality of the outgoing $e^{-}$ can be measured. The angular
resolution is expected to be $\Delta \theta \sim \Delta \phi \sim 30$
mrad $\sim 2^\circ$, substantially better than that of
Super-Kamiokande.  Notice also that the width of the Cerenkov cone
defined by the angle $\theta$ is very narrow for high-energy boron
neutrinos, as one can see from \eq{angles}.  In contrast, for $pp$
neutrino energies accessible at Hellaz ($T_{max} \simeq 0.26$ MeV,
$T_{th} \simeq 0.1$ MeV) we estimate that $\theta$ can be as large as
$48^\circ$.

Let us now discuss how the measurement of the azimuthal asymmetry could
be carried out considering that $\vec{B}_\odot$ is constant over a
given period of time but its direction is unknown. One should collect
events in every $\phi$-bin, where $\phi$ is defined with respect to
some arbitrarily chosen axis, and then take for different $\phi 's$
the ratio ${\cal A} (\phi ')$ which should show a $\sin \phi '$
dependence with a maximum equal to $A$.  This maximum will show us the
angle $\phi_{0}$ which corresponds to the direction of $\vec{B}_\odot$
($\phi_{0}=0$ if $\vec{B}_\odot$ goes along the positive x axis). Then
the direction of $\vec{B}_\odot$ is measured together with $A$. Since
this direction may change in time the experiment should accumulate
events until the maximum $\sin \phi $-like correlation in ${\cal A}
(\phi)$ is found and then start a new event counting period when such
correlation goes away due to the changing direction of
$\vec{B}_\odot$. Therefore the value of $A$ could be extracted by
performing a series of such measurements.
    
We have calculated the maximal integrated azimuthal asymmetry $A$ for
the case of $^8$B neutrinos, i.e.  for the situation described in
references \cite{Barbieri} and \cite{Vogel}. In order to do this we
have made use of the corresponding differential $^8$B neutrino
spectrum in the Standard Solar Model \cite{BP}.  Our results are shown
in figure \ref{asimhigh}. One can see that the asymmetry in the Dirac
diagonal case ($\nu_{e L} \to \nu_{e R}$) is in fact approximately a
factor two smaller than predicted by Barbieri \& Fiorentini. The
asymmetry in the Majorana case ($\nu_{e L} \to \bar{\nu}_{\mu R}$) for
the equivalent $\mu$ value is, as expected, smaller since there are
two active species in the neutrino flux so that the weak term (which
enters in the denominator in \eq{asym4}) becomes larger.

Figure \ref{asimlow} shows the results obtained by a similar analysis
for the case of $pp$-neutrinos, again taking into account the
theoretically predicted differential $pp$ neutrino spectrum and a more
realistic value of the survival probability for the RSFP scenario in
the Sun ($P_e=0.5$). Searching for such an asymmetry would be quite an
interesting physics task in an Hellaz-like experiment.

The dependence of $A$ on the value of the magnetic moment deserves a
more detailed analysis. One can write
\begin{equation}
A=\Frac{\mu_{\nu} a_{int}}{a_{weak}+\mu_{\nu}^2 a_{em}}
\end{equation}
with $\mu_{\nu} a_{int}=4n_{int}$, $a_{weak}=2\pi n_{weak}\equiv N_{weak}$ and
$\mu_{\nu}^{2} a_{em}=2\pi n_{em} \equiv N_{em}$. It follows that $A$ is
maximized for $N_{weak}=N_{em}$, e.g.  when the pure weak term is equal to
the electro-magnetic contribution. This fact favours $pp$-neutrinos
with respect to high energy neutrinos, since such a maximum is reached
for lower $\mu_{\nu}$ values precisely due to the lower energies
considered. In fact, for an energy threshold of recoil electrons
$W_{e}=0.1$ MeV (reachable at Hellaz) the maximal asymmetry is
reached for $\mu_{\nu}\simeq 3 \cdot 10^{-11}\mu_{B}$ in contrast to
Boron neutrinos, for which the maximal asymmetry is reached for
$\mu_{\nu} \simeq 10^{-10} \mu_{B}$. This way one sees that
figs.~\ref{asimhigh} and \ref{asimlow} describe approximately the most
favourable situation for measuring $A$ (see also figure \ref{asimmu}).

The fact that $dA/d\mu_{\nu}=0$ is reached when the number of weak
events is equal to the number of electro-magnetic events seems to
suggest that the measurement of the total number of events
$N_{weak}+N_{em}$ would be enough to rule out the values of $\mu_{\nu}$ to
which the asymmetry is sensitive. However, it is not so if one bears
in mind that:
\begin{enumerate}
\item
background events from other processes always increase the total
number of events. Thus one needs to perform a good subtraction of
background events to get some information on the $\mu_{\nu}$ term. On
the other hand, assuming that the background is isotropic in the
azimuthal plane, it should not be present in the numerator of $A$ and
then if some asymmetry is measured one can be confident that it is due
to a neutrino magnetic moment;
\item
the asymmetry is a ratio of event numbers. Thus global normalization
uncertainties (e.g. in the total neutrino fluxes) completely drop out
from the asymmetry. On the other hand energy-dependent uncertainties
(in the neutrino spectrum or detection efficiencies) will be reduced
since the same integrations appear in the numerator and in the
denominator;
\end{enumerate}

Last, but not least, if resonant spin flavour precessions really take
place, one can not use the standard weak cross section and subtract it
in order to get information on $\mu_{\nu}$; the signal becomes then
uncertain since the ``weak background'' becomes unknown. 

For all these reasons the asymmetry measurement is preferred. It
suffers, of course from the dependence on the magnetic field
direction, which is unknown. However, sensitivity to that information
is also of potential astrophysical interest.

\section{A simple model for the conversion probability}

We now consider how the above results are affected by the energy
dependence expected in the conversion probability in the RSFP
scenario. The general evolution Hamiltonian characterizing the system
of two Majorana neutrinos with a non-zero transition moment is
four-dimensional \cite{BFD}. For simplicity we illustrate the energy
dependence of the conversion probability in the simplest realization
of the RSFP scenario \cite{RFSP} where vacuum mixing is neglected. In
this case the Hamiltonian becomes $2 \times 2$ and the conversion
probability $P(\nu_{eL}\to  \bar{\nu}_{\mu R})$ is given
analytically as (see for instance \cite{KIM})
\begin{equation}
P (\nu_{e}\to  \bar{\nu}_{\mu })=
\frac{1}{2} - \left(\frac{1}{2} - P_{LZ}\right ) \cos 2\theta_m
\label{probconv}
\end{equation}
Here $\theta_m$ is the effective neutrino mixing angle in matter 
\begin{equation}
\tan 2\theta_m = \frac{2\mu_{12}B_\odot}
{\frac{\Delta m^2}{2\omega} - 2V_N - V_C}
\end{equation}
where $\mu_{12}$ denotes the transition magnetic moment, $\Delta m^2
=m_{\nu_\mu}^2 - m_{\nu_e}^2$ and $V_C$ ($V_N$) is the effective
potential of the neutrinos in the medium that arises due to charged
(neutral) current interactions. On the other hand $P_{LZ}$ is the
Landau-Zener transition probability,
\begin{equation}
P_{LZ} = \exp \left (-\frac{\pi}{4}Q\right ) \qquad \mbox{where}
\qquad Q \simeq \frac{16\omega (\mu_{12}B_\odot)^2 0.1 R_\odot}
{\Delta m^2}
\end{equation}
{}From analyses of experimental data and different models of the solar
magnetic field \cite{akhmedov97}, one finds that the data are well
reproduced for $\Delta m^2 \simeq 4 \cdot 10^{-9} - 2\cdot 10^{-8}$
eV$^2$, with a maximum $B_\odot$ strength of $25-50$ kG, assuming
$\mu_{12} = 10^{-11}\mu_B$.

For low-energy solar neutrinos, such as $pp$ neutrinos, the conversion
in \eq{probconv} corresponds to the non-adiabatic regime $Q \lsim 1$,
so that
\begin{equation}
P(\nu_{eL} \to  \bar{\nu}_{\mu R}) \simeq 1-P_{LZ} (\omega)
\label{probconv2}
\end{equation}
In fig.~\ref{PwithE} we show our results for the azimuthal asymmetry
of events when the conversion probability depends on the neutrino
energy as in \eq{probconv2}. We present three choices of $\Delta
m^2,B_\odot$ values that lead to a conversion probability of the order
of 40-60\%. The case of constant $P_e=0.5$ is shown for comparison.  One
can see from the figure that the dependence of $P_e$ on the neutrino
energy leads to somewhat smaller azimuthal asymmetries, but
qualitatively very similar to those obtained in the previous case,
where the energy dependence was neglected. There is basically a
compensation of the effect when integrating over energies with the
$\nu_{pp}$ spectrum, in such a way that this dependence is smoothed.

\section{Summary and Discussion}

Measuring azimuthal asymmetries in future low-energy solar
neutrino-electron scattering experiments with good angular resolution
should be a feasible and illuminating task.  Such asymmetries should
provide useful information on non-standard neutrino properties such as
magnetic moments, as well as on solar magnetic fields. The effect
follows from an interference of electro-magnetic and weak amplitudes
in the cross section.  We have seen that low-energy experiments such
as Hellaz (sensitive mainly to pp neutrinos) should provide a much
better means for the study of azimuthal asymmetries than accessible at
the Kamiokande or Super-Kamiokande experiments (sensitive to Boron
neutrinos).  For equal values of the magnetic moments, the expected
asymmetries are larger for Dirac neutrinos than for Majorana neutrino
transition moments. However, the Dirac neutrino case is probably less
likely, as there is no resonant conversion in the Sun. One exception
would be the case of Dirac neutrinos in the presence of twisting
magnetic fields \cite{akhpetsmi}. However, although in this case
resonant conversions in matter can take place one expects (as
mentioned in section 3) a washing out of the asymmetry effect due to
the changing magnetic field direction. Therefore the RSFP scenario
remains as the most promising possibility. It is also the most
interesting one theoretically, since Majorana neutrinos are more
fundamental and arise in most models of particle physics beyond the
Standard Model.

Note that the discussion given above we have assumed $\nu_e$ magnetic
moments of the order $10^{-11}\mu_B$ which is consistent with present
laboratory experiments. Apart from possible effects in red giants, a
$\nu_e$ transition moment of $10^{-11}\mu_B$ is compatible with
astrophysical limits, given the present uncertainties in these
considerations.

\section*{Acknowledgements}

The authors thank Tom Ypsilantis for fruitful discussions on the
Hellaz experiment.  This work has been supported by DGICYT under
Grants PB95-1077 and SAB95-506 (V.B.S.), by the TMR network grant
ERBFMRXCT960090 and by INTAS grant 96-0659 of the European
Union. S.P. was supported by Conselleria d'Educaci\'o i Ci\`encia of
Generalitat Valenciana. V.B.S. also acknowledges the support of RFFR
through grant 97-02-16501.

\newpage

\begin{figure}
\caption{Coordinate system conventions.}
\label{axis}
\end{figure}

\begin{figure}
\caption{Maximal integrated azimuthal asymmetry $A$ for Boron
neutrinos as a function of the electron recoil energy threshold
$W_e$. Solid line: Dirac case ($\nu_{e L} \to \nu_{eR}$).  Dashed
line: Majorana case ($\nu_{e } \to \bar{\nu}_{\mu }$). Here $P_e$ is
the electron neutrino survival probability.}
\label{asimhigh}
\end{figure}

\begin{figure}
\caption{Maximal integrated azimuthal asymmetry $A$ for $pp$ neutrinos
as a function of $W_e$.  Solid line: Dirac case. Dashed line: Majorana
case.}
\label{asimlow}
\end{figure}

\begin{figure}
\caption{Dependence of the maximal azimuthal symmetry A on the magnetic
moment.  Solid line: Dirac case. Dashed line: Majorana case. The
threshold energy for recoil electrons is fixed at $W_{e}=0.1$ MeV
(upper lines) and $W_{e}=0.05$ MeV (lower ones). }
\label{asimmu}
\end{figure}

\begin{figure}
\caption{The three lower curves show the maximal azimuthal asymmetry 
$A$ in the simple RSFP model described in section 4 for the Majorana
transition moment case for three different choices of parameters.  The
upper one gives the expected asymmetry expected for the case of
constant $P_e=0.5$ (maximum of asymmetry).}
\label{PwithE}
\end{figure}

\newpage
\thispagestyle{empty}
\centerline{\protect\hbox{\psfig{file=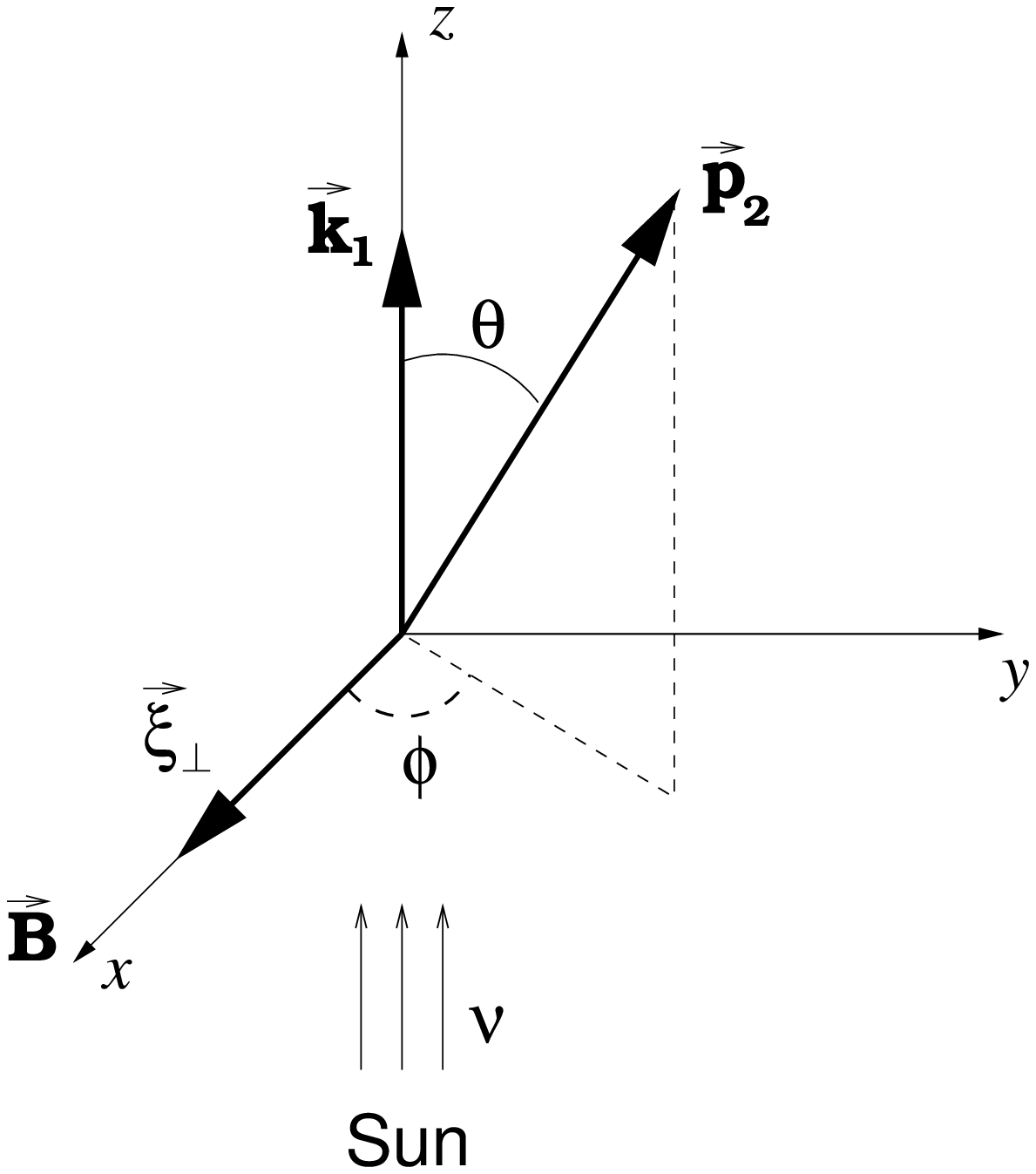,width=8cm}}}
Fig. 1\\
\newpage
\thispagestyle{empty}
\centerline{\protect\hbox{\psfig{file=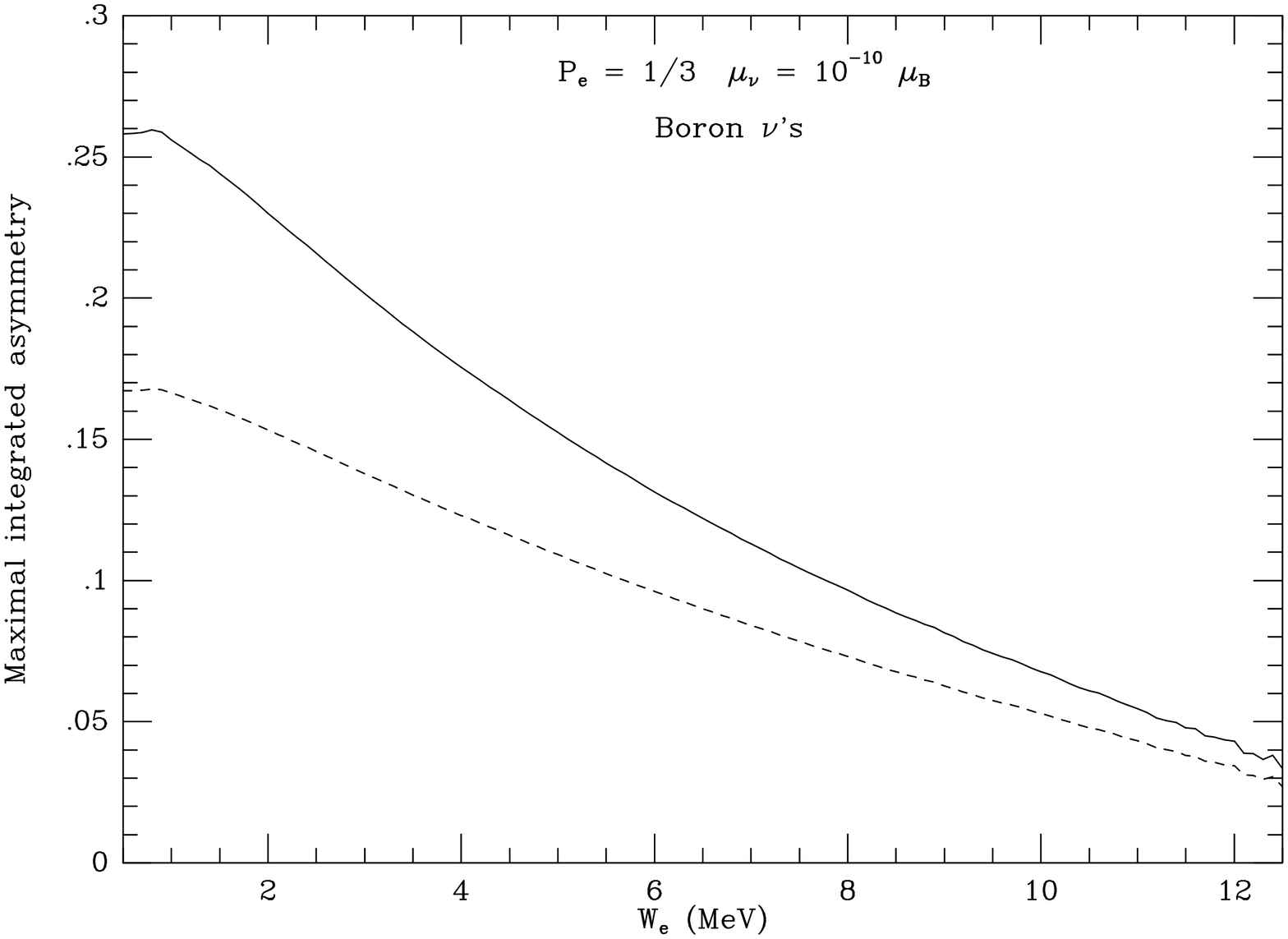,width=14cm,angle=-90}}}
Fig. 2\\
\newpage
\thispagestyle{empty}
\centerline{\protect\hbox{\psfig{file=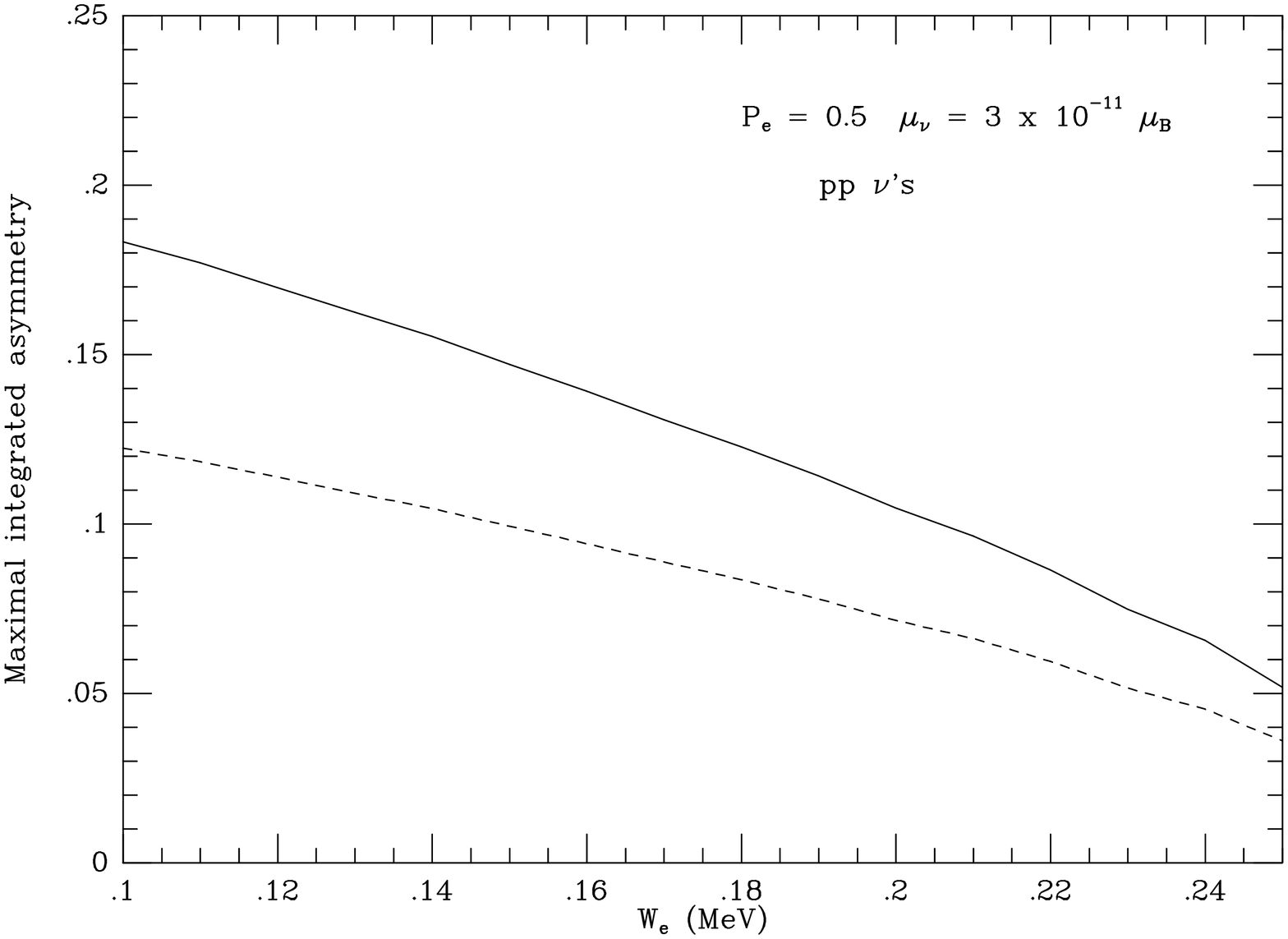,width=14cm,angle=-90}}}
Fig. 3\\
\newpage
\thispagestyle{empty}
\centerline{\protect\hbox{\psfig{file=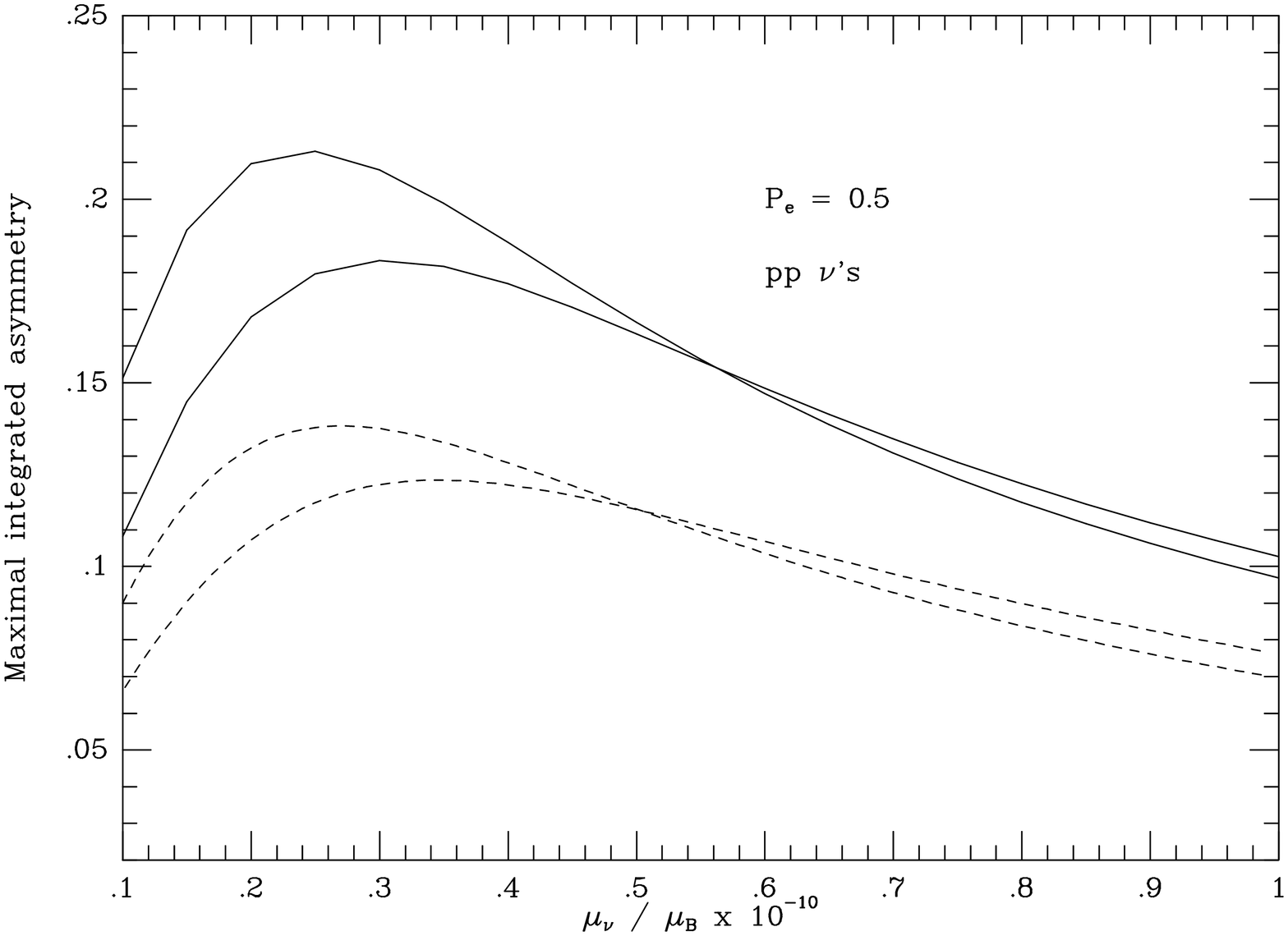,width=14cm,angle=-90}}}
Fig. 4\\
\newpage
\thispagestyle{empty}
\centerline{\protect\hbox{\psfig{file=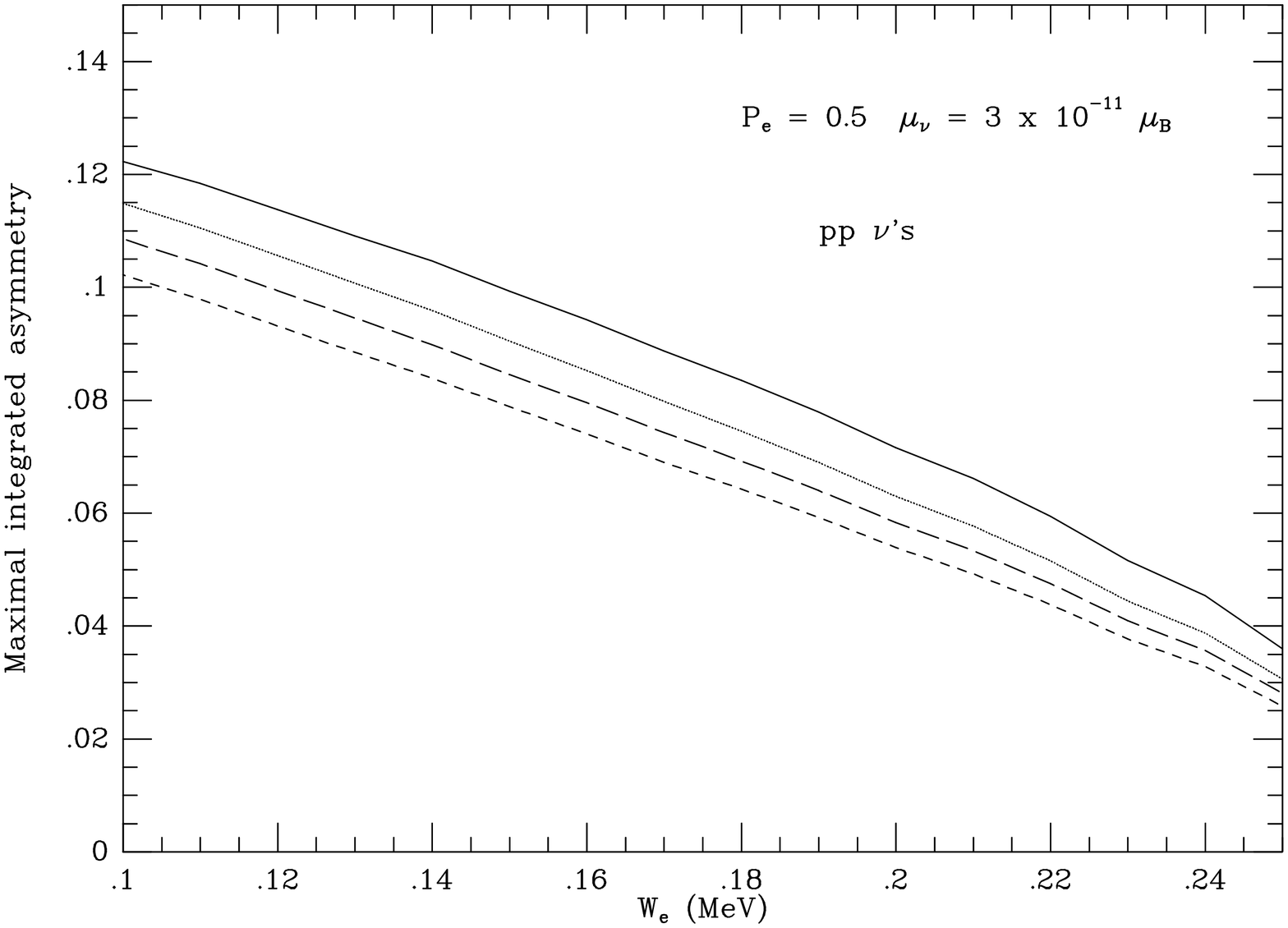,width=14cm,angle=-90}}}
Fig. 5\\
\end{document}